\documentclass[a4paper,noarxiv, twocolumn]{quantumarticle}
\usepackage[utf8]{inputenc}
\usepackage[english]{babel}
\usepackage[T1]{fontenc}
\usepackage[ruled,vlined]{algorithm2e}

\usepackage{hyperref}
\usepackage{breakurl}

\usepackage{amsmath}
\usepackage{amssymb}

\usepackage{centernot}
\usepackage{mathtools}
\usepackage{stmaryrd}

\makeatletter
\newcommand{\xMapsto}[2][]{\ext@arrow 0599{\Mapstofill@}{#1}{#2}}
\def\Mapstofill@{\arrowfill@{\Mapstochar\Relbar}\Relbar\Rightarrow}
\makeatother

\usepackage{geometry}
\usepackage{listings}
\usepackage{xcolor}

\definecolor{codegreen}{rgb}{0,0.6,0}
\definecolor{codegray}{rgb}{0.5,0.5,0.5}
\definecolor{codepurple}{rgb}{0.58,0,0.82}
\definecolor{backcolour}{rgb}{0.95,0.95,0.92}

\lstdefinestyle{mystyle}{
    backgroundcolor=\color{backcolour},
    commentstyle=\color{codegreen},
    keywordstyle=\color{magenta},
    numberstyle=\tiny\color{codegray},
    stringstyle=\color{codepurple},
    basicstyle=\ttfamily\footnotesize,
    breakatwhitespace=false,
    breaklines=true,
    captionpos=b,
    keepspaces=true,
    numbers=left,
    numbersep=5pt,
    showspaces=false,                
    showstringspaces=false,
    showtabs=false,
    tabsize=2
}

\title{Polynomial speedup in Torontonian calculation by a scalable recursive algorithm}
\author{Ágoston Kaposi}
\author{Zoltán Kolarovszki}
\author{Tamás Kozsik}
\affiliation{Department of Programming Languages and Compilers, E\"otv\"os  Lor\'and  University, Budapest, Hungary}
\author{Zoltán Zimborás}
\affiliation{Wigner Research Centre for Physics of the Hungarian Academy of Sciences, Budapest, Hungary}
\affiliation{BME-MTA Lendület Quantum Information Theory Research Group, Budapest, Hungary}
\author{Péter Rakyta}
\affiliation{Department of Physics of Complex Systems, E\"otv\"os  Lor\'and  University, Budapest, Hungary}

\begin{document}

\maketitle

\begin{abstract}

Evaluating the Torontonian function is a central computational challenge in the simulation of Gaussian Boson Sampling (GBS) with threshold detection. 
In this work, we propose a recursive algorithm providing a polynomial speedup in the exact calculation of the Torontonian compared to  state-of-the-art algorithms.
According to our numerical analysis the complexity of the algorithm is proportional to $N^{1.0691}2^{N/2}$ with $N$ being the size of the problem.
We also show that the recursive algorithm can be scaled up to HPC use cases making feasible the simulation of threshold GBS up to $35-40$ photon clicks without the needs of large-scale computational capacities.
\end{abstract}

\section{Introduction}


During the last two decades, promising steps were made towards constructing useful quantum computing hardware. In this endeavor, the first major intermediate goal is the demonstration of a scalable quantum advantage or quantum computational supremacy over classical computers \cite{preskill2012quantum, lund2017quantum, harrow2017quantum}. Among the different approaches for demonstrating quantum advantage  \cite{Aaronson:14, bremner2016average,  boixo2018characterizing, bouland2019complexity, haferkamp2020closing, oszmaniec2020fermion}, photonics provides a promising track as it enables fast gate operations,  room-temperature functioning and significant potential for scalability \cite{bourassa2021blueprint, bartolucci2021fusion,taballione2021universal}.  
The most well-known and feasible photonic advantage scheme is
\emph{Boson Sampling}  formulated by Aaronson and Arkhipov \cite{Aaronson:14} and its extended variant known as  \emph{Gaussian Boson Sampling} (GBS) \cite{PhysRevLett.119.170501,PhysRevA.100.032326,PhysRevA.98.062322}, which
are shown to be  classically intractable  under certain assumptions  \cite{Aaronson:14, deshpande2021quantum}. However, it should also be mentioned that high loss rates, partial photon indistinguishably and other issues could make such systems classically simulable \cite{oszmaniec2018classical,garcia2019simulating, qi2020regimes, renema2020simulability, brod2020classical, renema2020marginal}. 

On the experimental side, after the proof-of-principle realizations of small-scaled photonic interferometers \cite{Broome794,Spring798,Crespi2013,Tillmann2013}, more advanced devices were reported by increasing the number of modes on the photonic chip \cite{Spagnolo2014} and by multiplexing on-chip single-photon sources \cite{Spring:17,Faruque:18,doi:10.1116/5.0018594}.
An alternative approach to scale up Boson Sampling experiments was proposed in Refs.~\cite{PhysRevLett.113.100502,Bentivegnae1400255,PhysRevLett.118.020502} by sending multiple heralded single photons, shot by shot, into different random input ports of the interferometer.
In addition, several alternative schemes were proposed to overcome the state-preparation problem arising in photonic systems by putting the Boson Sampling problem into time domain \cite{PhysRevLett.113.120501,PhysRevA.93.043803}, or by the implementation of Boson Sampling with trapped ion technology \cite{PhysRevLett.112.050504,Toyoda2015} or with microwave cavities \cite{PhysRevLett.117.140505}.
In 2020 the GBS quantum computational advantage was demonstrated  via sampling from a Gaussian state at the output of a $100$-mode ultralow-loss interferometer with threshold detection \cite{Zhong1460} and an average of around $45$ photons. Subsequently this was extended to $144$ modes with programmable input states \cite{zhong2021phase}. 

In parallel to the development of experimental devices exhibiting verifiable quantum (Gaussian) Boson Sampling, efforts were also devoted to classically simulate increasingly larger GBS set-ups.
The classical simulation of these problems play an important role in their development. 
Without doubts, the ability to simulate larger quantum systems on smaller computers makes the studies of quantum algorithms a  lot easier.
Generally, the efficiency of quantum simulators can be traced down to well defined problems making the simulation classically intractable.
In case of GBS such central problem is the evaluation of the Hafnian \cite{PhysRevLett.119.170501,bulmer2021boundary} or the Torontonian \cite{PhysRevA.98.062322,DBLP:journals/corr/abs-2009-01177} function. 
In particular, the Torontonian can be used to calculate the output probabilities when threshold detection is used to sample the Gaussian state \cite{PhysRevA.98.062322,DBLP:journals/corr/abs-2009-01177}. 
This detection type can make difference whether there were $0$ photons or finite number of photons at a given output port of the interferometer. 
We note that recently a more efficient method was reported in Ref.~\cite{bulmer2021boundary} to simulate GBS with threshold detection.
However, in contrast with the sampling algorithm based on the Torontonian function,
this method does not provide the probabilities of the individual samples, which might be important for validation tests.

The input of the Torontonian function is a self-adjoint positive definite matrix (with eigenvalues less than unity)
\begin{equation} \label{eq:A}
    \mathbf{A} = \mathbf{I}-\mathbf{\Sigma}^{-1}
\end{equation}
acquired from the complex Husimi covariance matrix $\mathbf{\Sigma}$ describing the Gaussian state at the output of the interferometer before any detection \cite{PhysRevA.98.062322}.
Provided that the interferometer is equipped with $d$ output ports to be measured, the sampling matrix $\mathbf{A}$ would contain $N=2d$ rows and columns.
The Torontonian $\textrm{Tor}(\dot)$ of this matrix is defined by the expression \cite{DBLP:journals/corr/abs-2009-01177}:
\begin{equation}
 \textrm{Tor}(\mathbf{A}) = \sum\limits_{Z\in P_N}  \frac{(-1)^{N/2-|Z|}}{\sqrt{|\textrm{det}(\mathbf{I}-\mathbf{A_Z})|}}\;, \label{eq:tor}
\end{equation}
where $P_N$ is the power set of $1,2,\dots, N/2$, leading to the summation of $2^{N/2}$ determinant terms. 
The matrix $\mathbf{A_Z}$ is constructed from the rows and columns of matrix $\mathbf{A}$ according to the elements of $Z$: 
the elements of $Z$ labels the optical modes from which matrix $\mathbf{A_Z}$ is constructed.

Equation (\ref{eq:tor}) clearly indicates the complexity associated with the evaluation of the Torontonian function. 
The evaluation time of the exponentially large summation is heavily affected by the computational complexity of the determinant function.
Currently the best known algorithms to calculate the determinant of an $N\times N$ matrix scale with $\sim N^3$ arithmetic operations. 
Since Eq.~(\ref{eq:tor}) implies the elaboration of exponentially many variable sized matrices, the polynomial factor in the overall complexity of Eq.~(\ref{eq:tor}) would be somewhat less than $N^3$. 
As we will see in Sec.\ref{sec:complexity}, the complexity of Eq.~(\ref{eq:tor}) can be characterized by $\sim N^{\omega}2^{N/2}$, where numerical results yield $\omega=2.7396$.

In this paper we show that the $N^{\omega}$ polynomial factor of the computational complexity can be reduced to $N^{1.0691}$ by redesigning the evaluation order of the addends in the sum of Eq.~(\ref{eq:tor}) and reuse the intermediate results of the determinant calculation to compute subsequent addends.
Compared to the scalable structure of Eq.~(\ref{eq:tor}) this recursive strategy implies data dependency in the evaluation process of the Torontonian, that leads to the reduction of the initial scalability of the algorithm.
We will show that it is possible to implement the outlined recursive algorithm by an efficient task oriented parallel model enabling the computation of the Torontonian much faster than previous implementations.
Depending on the size of the input matrix, we achieved up to $200-300$-fold speedup on traditional multithreaded PCs and HPC cluster nodes compared to the implementation provided by the well known TheWalrus package \cite{Gupt2019}. We used TheWalrus version 0.15.
We also propose an adaptive \emph{Message Passing Interface} MPI \cite{10.5555/898758} implementation of the algorithm that can be used to distribute the Torontonian calculation over several nodes of an HPC.
The developed recursive Torontonian algorithm is going to be shipped as a part of a larger photonic quantum computer simulation package \cite{piquassoboost}. 

The rest of the paper is organised as follows: 
In Sec.~\ref{sec:recursive} we provide the detailed description of the recursive algorithm to calculate the Torontonian. 
Then in Secs.~\ref{sec:complexity} and \ref{sec:scalability} we numerically examine the computational complexity and scalability of the recursive algorithm and we compare it to previously used computational strategies.
In Sec.~\ref{sec:fidelity} we address the question of numerical fidelity of calculating the Torontonian, and we provide a parameter regime where the numerical value of the Torontonian is the most reliable.
Finally, in Sec.~\ref{sec:benchmark} we benchmark our implementations of calculating the Torontonian.
We conclude our work in Sec.~\ref{sec:conclusions}. 

\section{The recursive Torontonian algorithm} \label{sec:recursive}

As we mentioned in the introduction, the computational costs of evaluating the Torontonian function can be significantly reduced by a wise reuse strategy of the intermediate results during the evaluation of the determinants in Eq.~(\ref{eq:tor}).
In our algorithm we calculate determinants via Cholesky decomposition~\cite{fa13248b3e0a4fa6b7774c0a96d2551a} bringing an $N\times N$ positive definite Hermitian matrix $A$ into a product form $A=LL^{\dagger}$, where $L$ is a lower triangular matrix.
Then the determinant of $A$ can be calculated from the diagonal elements of $L$ by $\textrm{det}(A) = \prod\limits_{i=1}^N|L_{ii}|^2$.

In this article we propose an algorithm to calculate the Torontonian where the determinant calculation routines are called recursively providing a possibility to partially reuse the Cholesky decomposition of each submatrix $A_Z$ in the next step of the recursive chain.
Our algorithm was designed to take advantage from reusable computational data, while keeping up an efficient parallelization technique and polynomial memory needs.
\begin{algorithm}[!ht]
\SetAlgoLined
\SetKwInOut{Input}{input}\SetKwInOut{Output}{output}
\Input{Hermitian, positive definite matrix $\mathbf{A}$ of size $N\times N$}
\Output{Lower triangular matrix $\mathbf{L}$}
 \For{$i\gets1$ \KwTo $N$} {
 
   \For{$j\gets1$ \KwTo $i-1$} { 
   
     $\mathbf{L}_{i,j} = \frac{1}{\mathbf{L}_{j,j}} \left( \mathbf{A}_{i,j} - \sum\limits_{k=1}^{j-1} \mathbf{L}_{i,k}\overline{\mathbf{L}_{j,k}}  \right)$
   
   }
   
   $\mathbf{L}_{i,i} = \sqrt{ \mathbf{A}_{i,i} - \sum\limits_{k=1}^{i-1} |\mathbf{L}_{i,k}|^2  }$
 
 }
 \caption{Cholesky decomposition} \label{alg:cholesky_decomp}
\end{algorithm}

\begin{figure}[!ht]
\begin{centering}
\includegraphics[width = 0.47\textwidth]{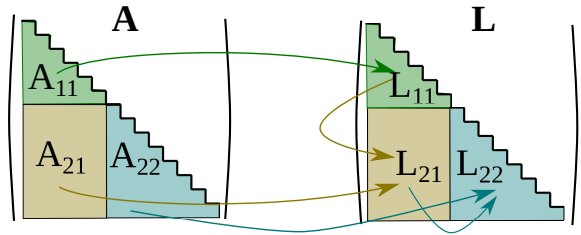}
\caption{Data dependency of the Cholesky decomposition. Sub-block $L_{11}$ depends only on $A_{11}$, $L_{21}$ can be calculated using $A_{21}$ and $L_{11}$, while $L_{22}$ is determined from elements of $A_{22}$ and $L_{21}$.}
\label{fig:cholesky_decomp}
\end{centering}
\end{figure}

\emph{Data dependency in Cholesky decomposition:} In order to explain the ideas behind our algorithm we briefly revisit the well-known algorithm of the Cholesky decomposition.
Algorithm \ref{alg:cholesky_decomp} shows the central formulas to determine the elements of the lower triangular matrix $L$ from an arbitrary $N\times N$ Hermitian and positive definite matrix $A$.
The decomposition follows a unique data dependency that is illustrated in Fig. \ref{fig:cholesky_decomp}.
Since $A$ is Hermitian and $L$ is lower triangular, it is sufficient to consider only the lower triangular part of both matrices.
The $L_{11}$ sub-block of the result matrix $L$ can be computed solely from the sub-matrix $A_{11}$.
To compute the sub-block $L_{21}$ wee need to know matrix elements of $A_{21}$ and $L_{11}$.
Finally, $L_{22}$ depends on the block $A_{22}$ of the input matrix and on $L_{21}$.

After understanding the data dependency of the Cholesky decomposition we can examine possible data reuse strategies when computing the decomposition of matrices created by the removal of several rows and columns from the initial matrix for which the Cholesky decomposition is known.
\begin{figure}[!ht]
\begin{centering}
\includegraphics[width = 0.47\textwidth]{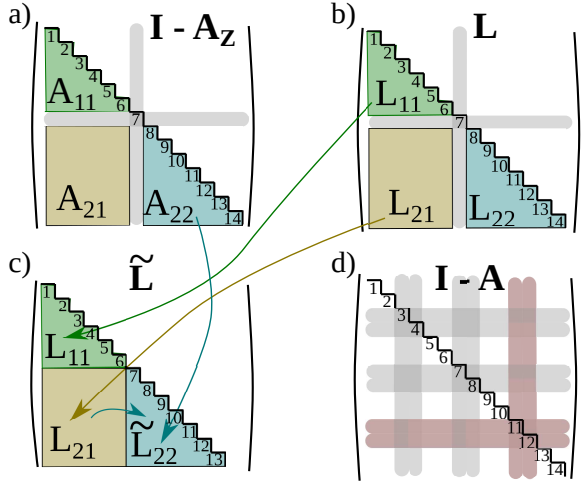}
\caption{a) The input matrix $\mathbf{I} - \mathbf{A_Z}$ in Eq.~(\ref{eq:tor}) from which a row and a column is removed in each recursive step to obtain the new Cholesky decomposition $\tilde{L}$. 
b) The Cholesky decomposition of matrix $\mathbf{I} - \mathbf{A_Z}$ of sub-figure a). 
The arrows pointing to sub-figure c) indicates data reuse to obtain the Cholesky decomposition of $\mathbf{I} - \mathbf{A_Z}$ from which the $7$-th rows and column were removed.
d) In each recursion step a pair of subsequent rows and columns is removed from the input matrix $\mathbf{I} - \mathbf{A}$ (indicated by the red areas). The row [column] to be removed from the matrix always has a greater row [column] index than the indices of previously removed rows [columns].}
\label{fig:remove_elements}
\end{centering}
\end{figure}
In order to illustrate the main concept behind the data reuse strategy, we assume a situation shown in Fig.~\ref{fig:remove_elements}. 
In this particular case we are about to remove the $7$th row and column from an initial matrix $\mathbf{A}$ of size $N=14$.
As indicated by the colored arrows pointing to Fig.~\ref{fig:remove_elements}.c the Cholesky decomposition $\mathbf{\widetilde{L}}$ of the reduced $13\times13$ matrix can be partially constructed from the elements of the Cholesky decomposition of the initial matrix, labeled by $\mathbf{L}$ in Fig.~\ref{fig:remove_elements}.b.
Namely, blocks $L_{11}$ and $L_{21}$ are identical in $\mathbf{L}$ and $\mathbf{\widetilde{L}}$.
Arithmetic operations are needed only to calculate the block $\widetilde{L}_{22}$ of $\mathbf{\widetilde{L}}$.
At this point the calculations involve the blocks $L_{21}$ and $A_{22}$ of the initial matrix.
As one can see, data reuse may significantly reduce the number of arithmetic operations when calculating the determinants of sub-matrices.

\emph{Data reuse in Torontonian calculation:}
Now we turn our attention to the structure of the recursive algorithm which tries to maximize data reuse during the evaluation of the Torontonian function.
We start the recursive algorithm by the calculation of the Cholesky decomposition of the matrix $\mathbf{A_Z} = \mathbf{A}$ containing all the optical modes of the input matrix.
In each step of the recursion  new matrices $\mathbf{A_Z}$ are constructed from the previous one by removing one optical mode.
The removal of one optical mode means deleting two columns and two rows from the matrix, since each optical mode is represented by two degrees of freedom.
Fig.~\ref{fig:remove_elements}.d illustrates the removal of one optical mode indicated by the red areas. 
(The optical modes corresponding to the grey-colored areas were removed in preceding recursion steps.)
In order to maximize the amount of reusable data, we consider the columns and rows of the input matrix $A$ to be ordered as {$\hat{a}_1$, $\hat{a}^{\dagger}_1$, $\hat{a}_2$,  $\hat{a}^{\dagger}_2$, $\dots$, $\hat{a}_{N/2}$, $\hat{a}^{\dagger}_{N/2}$}.
(Since the reordering of the input matrix implies even number of elementary permutations -- both rows and columns need to be permuted -- the Torontonian remains invariant under such operation.)
In each iteration we calculate the Cholesky decomposition of the constructed matrix $\mathbf{I} - \mathbf{A_Z}$ and derive one addend to the Torontonian of Eq.~(\ref{eq:tor}).
According to our reasoning in previous paragraphs, in each recursion step we can reuse the top left and the bottom left corners of the Cholesky decomposition matrix $\mathbf{L}$ calculated in the preceding iteration step (see blocks $L_{11}$ and $L_{21}$ in Fig.~\ref{fig:remove_elements}.b).

The structure of the recursive algorithm following the ideas outlined above is formulated in Alg. \ref{alg:recursive_alg}.
\begin{algorithm}[!ht]
\SetAlgoLined
\LinesNumbered
\SetKwFunction{FRecursiveTor}{RecursiveTor}

\SetKwInOut{Input}{input}\SetKwInOut{Output}{output}
\SetKwFor{For}{for}{:}{}
\SetKwIF{If}{ElseIf}{Else}{if}{:}{elif}{else:}{}
\SetKwFor{While}{while}{\{}{}

\Input{$\mathbf{A}$ an $N\times N$ sampling matrix defined by Eq.~(\ref{eq:A})}
\Output{The calculated Torontonian $\mathrm{tor}$}

 $\textbf{L}= \textrm{cholesky}(\mathbf{I} - \mathbf{A})$\;
 $\textrm{det}\left(\mathbf{I} - \mathbf{A}\right) = \prod\limits_{i=0}^{N-1} |\textbf{L}_{ii}|^2$\;
 $\textrm{tor} =  \frac{1}{\sqrt{|\textrm{det}(\mathbf{I} - \mathbf{A})|}}$\;

$\textrm{recursiveTor}(\textbf{L}, [])$\;

\SetKwFunction{FRecursiveTor}{recursiveTor}
\SetKwProg{Fn}{function}{:}{}
\Fn{\FRecursiveTor{$\mathbf{L}$, $\textrm{modes}$}}{
    \If{($\textrm{size}(\textrm{modes}) == 0$)} {
        $\textrm{start} = 0$\;
    }\Else {
        $\textrm{start} = \textrm{modes}[end] + 1$\;
    } 
    
    \For{($i=\textrm{start}$; $i<N/2$; $i$++)} { 
        $\textrm{nextModes} = \textrm{modes} + [i]$\;
        $\mathbf{A_Z} = \textrm{createAZ}(\mathbf{L}, \textrm{nextModes})$\;
        $\mathbf{\widetilde{L}} = \textrm{cholesky}(\mathbf{I} - \mathbf{A_Z})$\;
        
        $\textrm{determinant} = \prod\limits_{i=0}^{N-2|\textrm{nextModes}|-1} |\widetilde{L}_{ii}|^2$\;

        $\textrm{tor} = \textrm{tor} +  \frac{(-1)^{|\textrm{nextModes}|}}{\sqrt{\textrm{determinant}}}$;
        
        $\textrm{recursiveTor}(\mathbf{\widetilde{L}}, \textrm{nextModes})$\;
    }
    
        \KwRet\;
}
 
 \caption{Recursive Torontonian algorithm} \label{alg:recursive_alg}
\end{algorithm}
After calculating the determinant of the input matrix (line $2$ of Alg. \ref{alg:recursive_alg}) and initializing the Torontonian with the first addend (line $3$ of Alg. \ref{alg:recursive_alg}) a function call is made to start the recursive iterations.
The function \emph{recursiveTor} accepts two inputs: (i) the Cholesky Decomposition $\mathbf{L}$ reused from the previous iteration step (ii) and a list $modes$ of optical modes that were removed from the initial matrix when $\mathbf{L}$ was obtained in the preceding iteration cycle.
(In line $4$ of Alg. \ref{alg:recursive_alg} the iterations are started with an empty $modes$.)
As one can see, each recursion step involves a $\textrm{For}$ loop iterating over the optical modes that are appended to $modes$, and in the loop body a new -- partially decomposed -- $\mathbf{A_Z}$ matrix is constructed via removing one optical mode from $\mathbf{L}$ and by inserting matrix elements from $\mathbf{A}$ into sub-block $22$ which cannot be reused from $\mathbf{L}$ (see Fig.~\ref{fig:remove_elements}).
After finalizing the Cholesky decomposition of the newly created $\mathbf{A_Z}$ (in the implementation this is done in-place to avoid unnecessary memory operations) a new addend to the Torontonian can be calculated and new iterations can be spawned from line $16$ of Alg. \ref{alg:recursive_alg}.

In order to avoid duplicated calculations during the recursive calls, the new element added to  $modes$ in line $10$ of Alg. \ref{alg:recursive_alg} is always greater than the previous elements of $modes$.
The starting index $start$ of the $\textrm{For}$ loop is determined by the condition on lines $6-9$ of Alg. \ref{alg:recursive_alg}.
Consequently, the variable length of the $\textrm{For}$ loop varies during the evaluation process, making it somewhat harder to parallelize the algorithm. 
In Sec.\ref{sec:scalability} we will come back to this important question and we will present our numerical results on the scaling behaviour of our implementation.
Now we turn our attention to the complexity analysis of the recursive algorithm, and we compare it to the standard evaluation strategy, i.e.\ when each determinant in Eq.~(\ref{eq:tor}) is evaluated independently.

\section{Complexity of the recursive algorithm} \label{sec:complexity}

\begin{figure}[!ht]
\begin{centering}
\includegraphics[width = 0.47\textwidth]{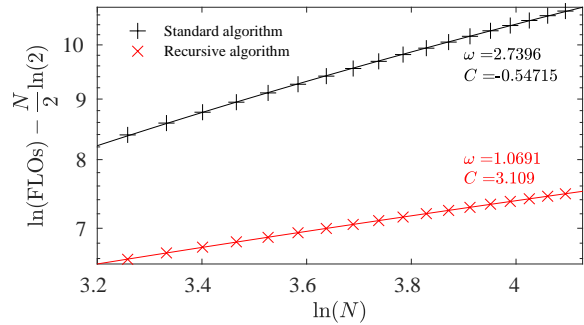}
\caption{Extracting the exponent $\omega$ of the polynomial complexity factor of the individual Torontonian calculation algorithms. The markers indicate the numerically obtained FLOs, and the solid lines represent the linear fit to the data sets. As stated in the main text, $\omega$ is given by the slope of the lines fitted to the $\textrm{ln}(\textrm{FLOs}) - \frac{N}{2} \textrm{ln}(2) - C$ curves, where $C$ and $\omega$ are the fitting parameters.}
\label{fig:torontonian_benchmark_omega}
\end{centering}
\end{figure}

Due to the high data-reuse tendency of the recursive algorithm the analytical estimation of its complexity is quite challenging.
Here we rather examine the Torontonian algorithms by a numerical procedure, i.e. instead of giving a closed formula to estimate the total number of floating point operations (FLOs) we sum up all the FLOs occurring during run-time by the code itself. 
This approach gives us the precise number of FLOs for both the standard and for the recursive computational strategy.
One might expect the complexity of the algorithms to be proportional to $N^\omega 2^{N/2}$. 
While the exponential factor originates from the exponentially large sum in Eq.~(\ref{eq:tor}), the polynomial factor $N^\omega$ comes from the underlying matrix operations.
The value of $\omega$ can be obtained from the logarithm of the overall FLOs as follows:
\begin{equation}
   \omega \textrm{ln}(N) = \textrm{ln}(\textrm{FLOs}) - \frac{N}{2} \textrm{ln}(2) - C\;, \label{eq:linear_fit}
\end{equation}
where $C$ is a constant number.
Consequently $\omega$ can be obtained by a simple linear fitting shown in  Fig.~\ref{fig:torontonian_benchmark_omega}, where we provide a comparison of two algorithms.
As one can see, the numerical data are in line with our expectation, the asymptotic scaling given by $\sim N^\omega 2^{N/2}$ is robust from matrix size $N=26$ [see $\textrm{ln}(26)=3.2581$ on the $x$ axis of Fig.~\ref{fig:torontonian_benchmark_omega}.].
For matrices smaller than $N=26$ the FLOs starts to deviate from the asymptotic linear relation of Eq.~(\ref{eq:linear_fit}). 
According to the numerical results, the recursive reuse of the Cholesky decomposition in subsequent determinant calculations reduces the computational complexity $\sim N^{2.7355}2^{N/2}$ of the standard algorithm to $\sim N^{1.0695}2^{N/2}$ leading to a polynomial speedup by a factor of $\sim N^{1.666}$. 
Thus, in case of an $N=50$ input matrix the recursive algorithm needs about $18$ times less arithmetic operations than the standard algorithm. 
For an input matrix of size $N=100$, the difference is even more striking, the standard algorithm requires $57$ times more arithmetic operations than the recursive algorithm.

\section{Scalability of the recursive algorithm} \label{sec:scalability}

In previous sections we constructed a recursive algorithm to evaluate the Torontonian function and we quantitatively examined its complexity compared to the standard algorithm.
However, from practical point of view the actual advantage of the recursive algorithm highly depends on its performance in parallel environments.
Since the calculation of the Torontonian scales exponentially with the size of the problem, it is inevitable to make use of modern hardware resources supporting different kinds of parallelism.
In this section we examine the scaling behaviour of the recursive algorithm implemented in two models of parallelism that are commonly used in HPC environments and are also available in standard PC's as well.

\begin{figure}[!ht]
\begin{centering}
\includegraphics[width = 0.47\textwidth]{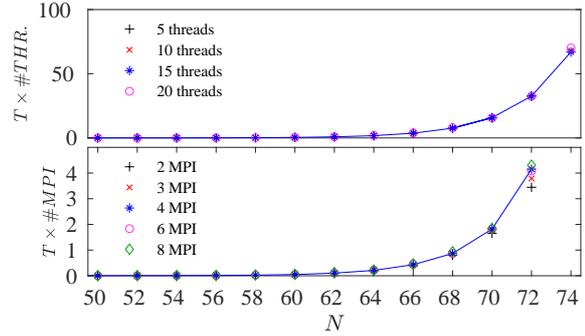}
\caption{a) The Torontonian computational time of a random sampling matrix in units of $10^4$ s multiplied by the number of threads used during the calculations on a single HPC node.
The data points corresponding to different number of threads fit onto a single line exhibiting a strong scaling of the underlaying shared memory parallelism. 
b) The Torontonian computational time of a random sampling matrix in units of $10^4$ s multiplied by the number of the MPI processes used in the calculations. 
Each MPI process is spawned over $10$ multi-core threads.
The data points of different number of MPI processes are close to fit onto a single line exhibiting a nearly strong scaling of the implementation over the involved HPC nodes.
The solid lines guide the eyes over the data points.
The calculations were performed on \emph{Xeon E5-2680v2 2.80GHz} platform.}
\label{fig:torontonian_benchmark_scaling}
\end{centering}
\end{figure}

In our algorithm each recursion step involves a $\textrm{For}$ loop (see line $8$ of  Alg. \ref{alg:recursive_alg}) spawning cycles that are independent from each other. 
Thus, a possible way to introduce parallelism into the algorithm is to execute the body of the  $\textrm{For}$ loop in parallel.
An apparent difficulty of this approach is related to the generation of nested parallelism, which poses strict limitations on the possible technical realization of the implementation.
To overcome this difficulty, we implemented the parallel version of the recursive algorithm via the Threading Building Block  (\href{https://github.com/oneapi-src/oneTBB}{TBB}) library \cite{ProTBB} providing an efficient task-oriented parallel programming interface to achieve an optimal workload balance among the accessible execution units of the underlying hardware in \emph{shared memory model} avoiding any over-subscription of the resources.
In addition, low level measures were taken to assure data correctness and race condition-free execution by introducing thread local storages to manipulate data structures. 
This way we managed to minimize the synchronization overhead between the threads, and we achieved ideal strong scaling of the implementation.
In order to illustrate the scaling properties of our implementation, we measured the execution time of the implementation to calculate the Torontonian on $5$, $10$, $15$ and $20$ threads. 
Figure \ref{fig:torontonian_benchmark_scaling}.a) shows the execution times multiplied by the corresponding number of the threads.
For implementations exhibiting strong scaling the data points are expected to collapse onto a single curve independently of the problem size.
According to our result shown in Fig.~\ref{fig:torontonian_benchmark_scaling}.a) our implementation is in line with this expectation providing a robust scalability in shared memory model of parallelism.

As one can see, the implementation of the recursive algorithm works well on centralized architectures like one node of an HPC cluster or a multicore workstation.
However, when it comes to large scaled problem, the distribution of the computational work over decentralized execution units becomes of high importance.
In the forthcoming paragraphs we demonstrate the possibility to scale up recursive Torontonian calculations via Message Processing Interface (MPI) \cite{10.5555/898758}.
Since the structure of the recursive algorithm is built on many variable sized $\textrm{For}$ loops nested into each other, the design of a balanced work distribution between the MPI processes seems to be quite challenging.
In order to overcome this issue we designed an adaptive MPI layer to support a dynamic work assignment between the allocated nodes of a standard HPC cluster.
\begin{figure}[!ht]
\begin{centering}
\includegraphics[width = 0.47\textwidth]{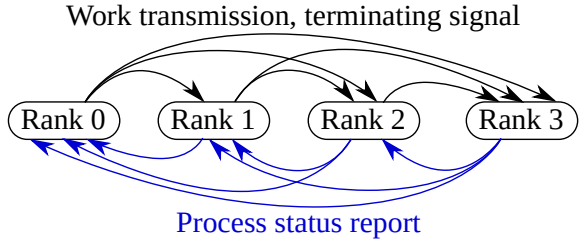}
\caption{The schematic design of the message passing strategy of the adaptive MPI layer between $4$ MPI processes. 
The MPI processes send to each other chunks of computational work, and reports back their current activity via status messages.
The MPI activity of one process is terminated by transmitting a terminating signal over the MPI network. For further explanation of the communication strategy see the main text.}
\label{fig:MPI_strategy}
\end{centering}
\end{figure}
Each MPI process is connected to the others by directed message paths.
Figure \ref{fig:MPI_strategy} shows an example consisting of $4$ MPI processes.
According to the designed MPI strategy, each process labeled by $\textrm{RANK}i$ can transmit work only to processes of higher rank, i.e. to $\textrm{RANK}j$'s, where $j>i$. 
When an MPI process $\textrm{RANK}i$ becomes idle or it starts to work on a received work, it immediately broadcasts a status message to indicate its ability to receive new work, or to inform the other processes to stop transmitting new work towards the current process.
The status message from $\textrm{RANK}i$ is transmitted only to processes of lower rank, namely for $\textrm{RANK}j$'s, where $j<i$.
As soon as a process finishes its work, it sends a terminating signal over channels used for work transmission.
An MPI process is allowed to finish its activity only in case it already received a terminating signal from all lower ranked processes.
Until the process does not receive all the necessary terminating signals, it keeps up to listen for the transmission of new assigned work.
When the MPI process is allowed to finish its activity, it broadcasts a terminating signal to all higher ranked MPI processes.
This way all MPI processes can finish their activity in a controlled way.

One might notice that the designed MPI communication strategy shown in Fig.~\ref{fig:MPI_strategy} is highly asymmetric, since processes are allowed to send computational work only to higher ranked processes.
Fortunately, this asymmetry fits well to the peculiar structure of the recursive algorithm.
For simplicity, let us consider only the very first $\textrm{For}$ loop in the recursive evaluation process at line $8$ of Alg. \ref{alg:recursive_alg}.
The computational time of the individual cycles spawned by the $\textrm{For}$ loop depends on the associated index $i$:
each cycle $i$ implies the calculation of $2^{N/2-i-1}$ addends to the Torontonian.
Consequently, an adequate initial work distribution is to divide the cycles of the first $\textrm{For}$ loop between the MPI processes, such that lower ranked processes are assigned to lower values of cycle index $i$.
This way lower ranked processes would always have more initial work to deal with, and they would be capable to transmit work to higher ranked process when the latter ones become idle.

Since the computational work to be transmitted can be fully characterized by the elements of the data structure $nextModes$ (see line $11$ of Alg. \ref{alg:recursive_alg}.), it is sufficient to transmit only the content of $nextModes$.
The receiving MPI process then initiates the recursive call on line $16$ in Alg. \ref{alg:recursive_alg} and continues the calculations taking over the place of the sender process.
(We also notice that transmitting the matrix $\widetilde{L}$ alongside $nextModes$ would cause large overhead in communication, so instead of transmitting the matrix $\widetilde{L}$, it is rather recalculated by the receiving process before it initiates the recursive call.)

In Fig.~\ref{fig:torontonian_benchmark_scaling}.b we show the outcome of our analysis.
We measured the calculation time of the Torontonian as a function of the problem size and multiplied the computational time by the number of the allocated MPI processes.
The measured data shown for the cases of $2$, $3$, $4$, $6$ and $8$ MPI processes are close to fall onto a single curve independently of the problem size $N$.
Thus, similarly to the case of the shared memory model studied earlier, the MPI implementation also exhibits a well-balanced workload among the allocated resources.
However, for larger matrices we can see a moderate divergence in the data points.
In particular, at problem size $N=72$ the computational time on $8$ MPI processes is only $3.2$ times faster than on $2$ MPI processes, while in case of perfect strong scaling the speedup factor would be exactly $4$.
This finding also indicates that the demonstrated scaling properties of the algorithm would not hold on up to measures used in Ref.~\cite{DBLP:journals/corr/abs-2009-01177}. 
Since our MPI model depends on active connections between all the involved nodes, spreading the calculations over several thousands of HPC nodes would result in significant communication overhead, killing the performance improvement coming from the recursive algorithm.

In summary, our scaling analysis showed that it is possible to bring the constructed recursive algorithm into HPC environments. 
Even if the algorithm itself does not scale well to large computing capacities, the main target of our efforts is the simulation of threshold GBS, which, on the other hand, can be scaled very well to large measures. 
As we will see by benchmark comparison discussed in Sec.~\ref{sec:benchmark}, using the developed recursive algorithm one can speed up the GBS simulation at least by two orders of magnitude.

Finally, let us formulate the concept of the parallelism behind our recursive implementation in a different form.
From bird's perspective we can think about the outlined task-oriented parallel models as a flow graph where the underlying data dependency specifies the direction of data flow between the individual nodes of the graph. 
The parallel capabilities of this model comes from the concurrent execution possibilities of the work chunks associated with the graph nodes. 
The computational work is picked up by the threads (or by MPI processes) dynamically, providing a well-balanced work distribution among the resources.
We remark that this parallel model is different from the data parallelism paradigm typically used to distribute work over high performance CPU and GPU clusters.
We believe that from application point of view, our model is most suitable for architectures supporting data flow type parallelism~\cite{Sutherland1966TheOG} such as FPGAs.

\section{Fidelity of Torontonian calculation} \label{sec:fidelity}

For practical use cases we need to consider another important aspect of the Torontonian calculators, namely their numerical fidelity. 
Here we note that the significant reduction of FLOs in the recursive algorithm is not accompanied with better numerical precision: while the algorithm reuses the intermediate results during the calculations, in the end the same number of arithmetic operations are used to obtain the individual addends to the Torontonian as in the standard algorithm. 
Consequently, from numerical precision point of view, the two algorithms have the same properties.

\begin{figure}[!ht]
\begin{centering}
\includegraphics[width = 0.47\textwidth]{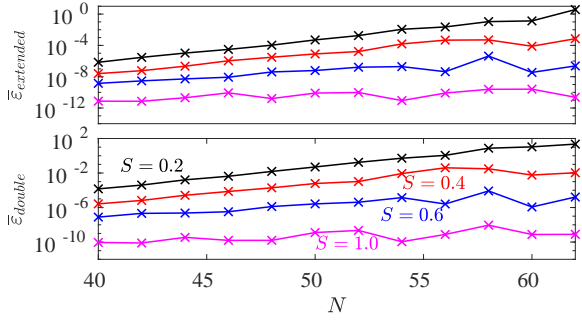}
\caption{The relative difference  of the calculated Torontonians between TheWalrus package and our implementation. 
The relative difference was averaged over the outcomes of $100$ and $10$ input matrices of size $N=40\dots52$ and $N=54\dots62$, respectively. 
$\varepsilon_X$ calculated for various squeezing parameters $S$ are shown for two cases:
in subfigure a) the calculations were carried out in extended precision in our implementation, while in subfigure b) we used double precision in the Cholesky decompositions.
The Torontonian calculations performed by the TheWalrus package were done in extended precision in both cases. The benchmark was done on an \emph{Intel Xeon Gold 6130} platform.}
\label{fig:torontonian_benchmark_error}
\end{centering}
\end{figure}

To get better insight into the numerical fidelity of the Torontonian calculation, we compared our implementation to a well known software tool developed by Xanadu.
TheWalrus package~\cite{Gupt2019} is part of the  Strawberry Fields \cite{strawberryfields} bosonic quantum computer simulation platform dedicated for various types of boson sampling simulations.
In Fig.~\ref{fig:torontonian_benchmark_error}. we plotted the relative difference
\begin{equation}
    \varepsilon_X = \frac{\left|\textrm{Tor}_{TheWalrus} - \textrm{Tor}_X\right|}{\textrm{Tor}_X}
\end{equation}
between the results of TheWalrus package and our implementation, where $\textrm{Tor}_{TheWalrus}$ labels the Torontonian calculated by TheWalrus package and $\textrm{Tor}_{X}$ stands for the Torontonian calculated by our implementation. 
We examined two levels of numerical precision provided by our implementation:
in the first case all calculations are done in extended precision ($X=extended$), while in the second case the Cholesky decomposition cycles are performed in double precision ($X=double$).
[In the later case the actual calculation of the determinants and the summation of Eq.~(\ref{eq:tor}) is still kept in extended precision.]
TheWalrus package calculates the Torontonian in extended precision.

Basically, the implementations studied in this section differs from each other only in the execution order of the arithmetic operations.
Hence, ideally they should give the same result, unless the calculated value of the Torontonian shows a dependence on the execution order of the arithmetic operations.
Due to the floating point representation of real numbers, different execution order of the code may result in different result, and it is not possible to mark one result to be more precise than another.
On the other hand, we can trust the calculated Torontonian if the results of the individual implementation are close to each other, since in this case they do not show dependence on numerical artifacts.

Following the reasoning of Ref.~\cite{DBLP:journals/corr/abs-2009-01177}, in realistic scenarios the Torontonian becomes a very small number. 
According to our numerical analysis, this happens when the average squeezing parameter characterizing the photonic modes is also a small number.
In this case, one needs to sum up exponentially many reciprocal square roots of determinants that are relatively close to unity in their magnitude.
To obtain the precise value of the Torontonian in such situation, one need to cast the calculations -- at least -- into extended precision. 
For example, when $0.2$ is set for the average squeezing parameter of the Gaussian state, the relative difference $\varepsilon_{extended}$ exceeds $1\%$ for matrices of size $N=62$, indicating that even extended precision is not good enough.
On top of that, when the Cholesky decomposition is performed in double precision, the relative difference $\varepsilon_{double}$ becomes even larger than $25\%$ at problem size $N=62$.

On the other hand, Torontonian calculations becomes much more reliable when the average squeezing parameter of the Gaussian state is larger.
If the squeezing parameter is set to $\sim1$ in average, the Torontonian becomes larger than one (instead of being a small value) and higher order decimals of the determinants becomes less important.
In such situation the relative difference $\varepsilon_{X}$ is less than $10^{-9}\%$ even if the Cholesky decomposition is calculated only in double precision.
The origin of this peculiar numerical effect can be traced back to the off-diagonal elements of the covariance matrix $\Sigma$ which are driven by the squeezing parameters of the optical modes.
When the average squeezing of the Gaussian state is low, the covariance matrix contains elements that differ from each other by many orders in magnitude.
The small off-diagonal elements contribute to the determinants by small corrections which tend to be of high importance when it comes to the summation of the addends in Eq.~(\ref{eq:tor}).
On the other hand, the elements of the covariance matrix become more balanced by increasing the average squeezing of the state and the Torontonian computation becomes numerically more stable. 
In Fig.~\ref{fig:torontonian_benchmark_error}. we plotted the relative difference $\varepsilon_X$ for various squeezing parameters. 
One can clearly see a crossover in the numerical stability when the average squeezing parameter $S$ is changed from $0.2$ up to $S=1$.
The Torontonian calculation is the less reliable when the squeezing parameter is low, while at $S=1$ even double precision is sufficient to calculate the Torontonian with a high fidelity.

We note that the evaluated Torontonian becomes again less reliable when $S$ further increases above unity. 
In this case the elements of the covariance matrix start to differ from each other by several orders of magnitude, and higher numeric precision would be required to calculate the Torontonian.


\section{Performance benchmark} \label{sec:benchmark}

In this section we provide a performance benchmark of our implementation based on the recursive Torontonian algorithm.
As in the previous section, we have chosen to compare our code to TheWalrus package, because the two implementations have many things in common: both of them are equipped with high level \emph{Python} interface to ease their usage, while beneath the surface the hard computational tasks are left for \emph{C/C++} engines.
This way we can compare implementations written in the same language and compiled on the same platform with the same compiler.
\begin{figure}[!ht]
\begin{centering}
\includegraphics[width = 0.47\textwidth]{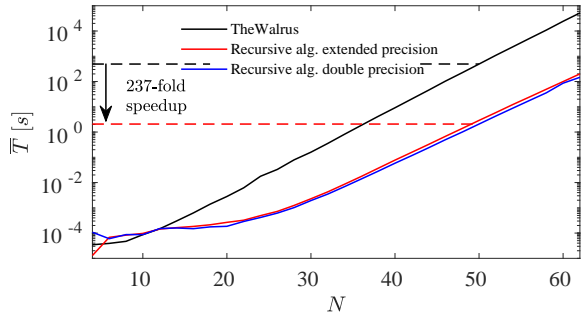}
\caption{Benchmark comparison of averaged Torontonian calculation time $\overline{T}$ of random sampling matrices defined by Eq. (\ref{eq:A}) as a function of the matrix size $N$.
As discussed in the main text, our implementation of the recursive algorithm shows much better performance then TheWalrus package version 0.15. 
The benchmark was done on an \emph{Intel Xeon Gold 6130} platform using $24$ threads in shared memory model. $\overline{T}$ was determined by the average time of $100$, $10$ and $2$ independent Torontonian computing cycles for matrices of size $N = 4\dots52$, $N=54\dots58$ and $N=60\dots62$ respectively.}
\label{fig:torontonian_benchmark}
\end{centering}
\end{figure}
While putting efforts to implement the most favourable algorithm to reduce the number of floating point operations, one must not forget about another important aspect of good run-time performance.
In order to reduce the computational time as much as possible we designed the structure of the implementation to keep the number of memory operations low, as well to reuse cache-line level data whenever it is possible.
The register-level parallelism via portable SIMD instructions is provided by the incorporation of low level BLAS and self-written AVX/AVX2 and AVX512F kernels. These kernels provide significant speedup in calculations involving double precision floating point representation, while in case of extended precision calculations (for example in some parts of the Hafnian calculators) such vectorization possibility is not supported by the hardware.

Figure \ref{fig:torontonian_benchmark}. compares the average Torontonian calculation time $\overline{T}$ of our implementation to TheWalrus code. 
We measured the computation time for two cases. 
In the first case (red curve in Fig.~\ref{fig:torontonian_benchmark}.) the calculations in our code are performed in extended precision, while in the second case the Cholesky decomposition is calculated in double precision (blue curve in Fig.~\ref{fig:torontonian_benchmark}.).
(According to our analysis of numerical fidelity, the latter case is also relevant for higher squeezing parameters.)
The performance difference between the two levels of numerical precision is in line with the expectation: the calculations in the double precision version are about $1.3-2$ times faster depending on the underlying platform and on the problem size due to AVX vectorization possibilities.

However, the performance difference compared to TheWalrus package is much more striking.
To emphasize the difference between the two software packages, we highlighted the speedup of our code for the problem size of $N=50$. 
In this particular case our code can calculate the Torontonian about $237$ times faster than TheWalrus package.
The remarkable speedup originates from two sources. 
As we estimated earlier in Sec.~\ref{sec:complexity}, from algorithmic point of view one can realize an about $\sim 18$-fold speedup at this specific problem size.
The remaining speedup factors come from the highly optimized code structure of our implementation.


The performance of the two implementations come close to each other only at small problem sizes.
This happens because the OpenMP library working behind TheWalrus package is more agile in acquiring hardware resources than the task stealing mechanism of the TBB library. 
At problem sizes less than $N=20$ our implementation effectively utilizes less hardware resources than we allocated for the benchmark comparison.
In this transition regime the calculation time of the Torontonian can be measured in milliseconds (or less).

\section{Conclusions and Outlook} \label{sec:conclusions}

The mathematical description of boson sampling simulations, including their numerically most challenging parts to evaluate the permanent, the Hafnian or the Torontonian functions imply the recalculation of many intermediate results over and over again.
In this work we developed a novel numerical approach to reuse already computed data during the calculation of the Torontonian, and have successfully reduced its computational complexity by a polynomial factor of $\sim N^{1.666}$.
The key ingredient of our approach relies on the peculiar structure of the Cholesky decomposition making the reuse of intermediate computational results in subsequent iteration cycles possible.
Efficient task-oriented parallel programming models enabled us to scale up the developed recursive algorithm to HPC measures. We also examined the numerical fidelity of Torontonian calculations.
In particular, we showed that Torontonian calculators reach their maximal fidelity at average squeezing parameter $S=1$.
Above and below of this limiting case, the Torontonian becomes numerically harder to calculate, needing to increase the numerical precision used in the computations.

Our benchmark comparisons with the state of the art implementation of TheWalrus package showed a significant speedup in the simulation of GBS with threshold detection.
The reduction of the execution time by two orders of magnitude makes it possible to simulate threshold GBS involving $40-50$ optical modes without relying on large-scale HPC resources.

\section{Acknowledgements}

The research was supported by the Ministry of Innovation and Technology and the National Research, Development and Innovation
Office within the Quantum Information National Laboratory of Hungary, and was
also supported by NKFIH through the Quantum Technology National Excellence Program
(No.2017-1.2.1-NKP-2017-00001) and  Grants No. K124152, FK135220, KH129601.
We acknowledge the computational resources provided by the Wigner Scientific Computational Laboratory (WSCLAB) (the formerWigner GPU Laboratory).

\bibliographystyle{unsrt}

\bibliography{bibliography}

\begin{thebibliography}{10}

\bibitem{preskill2012quantum}
John Preskill.
\newblock Quantum computing and the entanglement frontier, 2012.

\bibitem{lund2017quantum}
Austin~P Lund, Michael~J Bremner, and Timothy~C Ralph.
\newblock Quantum sampling problems, bosonsampling and quantum supremacy.
\newblock {\em npj Quantum Information}, 3(1):1--8, 2017.

\bibitem{harrow2017quantum}
Aram~W Harrow and Ashley Montanaro.
\newblock Quantum computational supremacy.
\newblock {\em Nature}, 549(7671):203--209, 2017.

\bibitem{Aaronson:14}
Scott Aaronson and Alex Arkhipov.
\newblock The computational complexity of linear optics.
\newblock In {\em Research in Optical Sciences}, page QTh1A.2. Optical Society
  of America, 2014.

\bibitem{bremner2016average}
Michael~J Bremner, Ashley Montanaro, and Dan~J Shepherd.
\newblock Average-case complexity versus approximate simulation of commuting
  quantum computations.
\newblock {\em Physical review letters}, 117(8):080501, 2016.

\bibitem{boixo2018characterizing}
Sergio Boixo, Sergei~V Isakov, Vadim~N Smelyanskiy, Ryan Babbush, Nan Ding,
  Zhang Jiang, Michael~J Bremner, John~M Martinis, and Hartmut Neven.
\newblock Characterizing quantum supremacy in near-term devices.
\newblock {\em Nature Physics}, 14(6):595--600, 2018.

\bibitem{bouland2019complexity}
Adam Bouland, Bill Fefferman, Chinmay Nirkhe, and Umesh Vazirani.
\newblock On the complexity and verification of quantum random circuit
  sampling.
\newblock {\em Nature Physics}, 15(2):159--163, 2019.

\bibitem{haferkamp2020closing}
Jonas Haferkamp, Dominik Hangleiter, Adam Bouland, Bill Fefferman, Jens Eisert,
  and Juani Bermejo-Vega.
\newblock Closing gaps of a quantum advantage with short-time hamiltonian
  dynamics.
\newblock {\em Physical Review Letters}, 125(25):250501, 2020.

\bibitem{oszmaniec2020fermion}
Micha{\l} Oszmaniec, Ninnat Dangniam, Mauro~ES Morales, and Zolt{\'a}n
  Zimbor{\'a}s.
\newblock Fermion sampling: a robust quantum computational advantage scheme
  using fermionic linear optics and magic input states.
\newblock {\em arXiv preprint arXiv:2012.15825}, 2020.

\bibitem{bourassa2021blueprint}
J~Eli Bourassa, Rafael~N Alexander, Michael Vasmer, Ashlesha Patil, Ilan
  Tzitrin, Takaya Matsuura, Daiqin Su, Ben~Q Baragiola, Saikat Guha, Guillaume
  Dauphinais, et~al.
\newblock Blueprint for a scalable photonic fault-tolerant quantum computer.
\newblock {\em Quantum}, 5:392, 2021.

\bibitem{bartolucci2021fusion}
Sara Bartolucci, Patrick Birchall, Hector Bombin, Hugo Cable, Chris Dawson,
  Mercedes Gimeno-Segovia, Eric Johnston, Konrad Kieling, Naomi Nickerson,
  Mihir Pant, et~al.
\newblock Fusion-based quantum computation.
\newblock {\em arXiv preprint arXiv:2101.09310}, 2021.

\bibitem{taballione2021universal}
Caterina Taballione, Reinier van~der Meer, Henk~J Snijders, Peter Hooijschuur,
  J{\"o}rn~P Epping, Michiel de~Goede, Ben Kassenberg, Pim Venderbosch, Chris
  Toebes, Hans van~den Vlekkert, et~al.
\newblock A universal fully reconfigurable 12-mode quantum photonic processor.
\newblock {\em Materials for Quantum Technology}, 1(3):035002, 2021.

\bibitem{PhysRevLett.119.170501}
Craig~S. Hamilton, Regina Kruse, Linda Sansoni, Sonja Barkhofen, Christine
  Silberhorn, and Igor Jex.
\newblock Gaussian boson sampling.
\newblock {\em Phys. Rev. Lett.}, 119:170501, Oct 2017.

\bibitem{PhysRevA.100.032326}
Regina Kruse, Craig~S. Hamilton, Linda Sansoni, Sonja Barkhofen, Christine
  Silberhorn, and Igor Jex.
\newblock Detailed study of gaussian boson sampling.
\newblock {\em Phys. Rev. A}, 100:032326, Sep 2019.

\bibitem{PhysRevA.98.062322}
Nicol\'as Quesada, Juan~Miguel Arrazola, and Nathan Killoran.
\newblock Gaussian boson sampling using threshold detectors.
\newblock {\em Phys. Rev. A}, 98:062322, Dec 2018.

\bibitem{deshpande2021quantum}
Abhinav Deshpande, Arthur Mehta, Trevor Vincent, Nicolas Quesada, Marcel
  Hinsche, Marios Ioannou, Lars Madsen, Jonathan Lavoie, Haoyu Qi, Jens Eisert,
  et~al.
\newblock Quantum computational supremacy via high-dimensional gaussian boson
  sampling.
\newblock {\em arXiv preprint arXiv:2102.12474}, 2021.

\bibitem{oszmaniec2018classical}
Micha{\l} Oszmaniec and Daniel~J Brod.
\newblock Classical simulation of photonic linear optics with lost particles.
\newblock {\em New Journal of Physics}, 20(9):092002, 2018.

\bibitem{garcia2019simulating}
Ra{\'u}l Garc{\'\i}a-Patr{\'o}n, Jelmer~J Renema, and Valery Shchesnovich.
\newblock Simulating boson sampling in lossy architectures.
\newblock {\em Quantum}, 3:169, 2019.

\bibitem{qi2020regimes}
Haoyu Qi, Daniel~J Brod, Nicol{\'a}s Quesada, and Ra{\'u}l
  Garc{\'\i}a-Patr{\'o}n.
\newblock Regimes of classical simulability for noisy gaussian boson sampling.
\newblock {\em Physical review letters}, 124(10):100502, 2020.

\bibitem{renema2020simulability}
Jelmer~J Renema.
\newblock Simulability of partially distinguishable superposition and gaussian
  boson sampling.
\newblock {\em Physical Review A}, 101(6):063840, 2020.

\bibitem{brod2020classical}
Daniel~Jost Brod and Micha{\l} Oszmaniec.
\newblock Classical simulation of linear optics subject to nonuniform losses.
\newblock {\em Quantum}, 4:267, 2020.

\bibitem{renema2020marginal}
Jelmer~J Renema.
\newblock Marginal probabilities in boson samplers with arbitrary input states.
\newblock {\em arXiv preprint arXiv:2012.14917}, 2020.

\bibitem{Broome794}
Matthew~A. Broome, Alessandro Fedrizzi, Saleh Rahimi-Keshari, Justin Dove,
  Scott Aaronson, Timothy~C. Ralph, and Andrew~G. White.
\newblock Photonic boson sampling in a tunable circuit.
\newblock {\em Science}, 339(6121):794--798, 2013.

\bibitem{Spring798}
Justin~B. Spring, Benjamin~J. Metcalf, Peter~C. Humphreys, W.~Steven
  Kolthammer, Xian-Min Jin, Marco Barbieri, Animesh Datta, Nicholas
  Thomas-Peter, Nathan~K. Langford, Dmytro Kundys, James~C. Gates, Brian~J.
  Smith, Peter G.~R. Smith, and Ian~A. Walmsley.
\newblock Boson sampling on a photonic chip.
\newblock {\em Science}, 339(6121):798--801, 2013.

\bibitem{Crespi2013}
Andrea Crespi, Roberto Osellame, Roberta Ramponi, Daniel~J. Brod, Ernesto~F.
  Galv{\~a}o, Nicol{\`o} Spagnolo, Chiara Vitelli, Enrico Maiorino, Paolo
  Mataloni, and Fabio Sciarrino.
\newblock Integrated multimode interferometers with arbitrary designs for
  photonic boson sampling.
\newblock {\em Nature Photonics}, 7(7):545--549, Jul 2013.

\bibitem{Tillmann2013}
Max Tillmann, Borivoje Daki{\'{c}}, Ren{\'e} Heilmann, Stefan Nolte, Alexander
  Szameit, and Philip Walther.
\newblock Experimental boson sampling.
\newblock {\em Nature Photonics}, 7(7):540--544, Jul 2013.

\bibitem{Spagnolo2014}
Nicol{\`o} Spagnolo, Chiara Vitelli, Marco Bentivegna, Daniel~J. Brod, Andrea
  Crespi, Fulvio Flamini, Sandro Giacomini, Giorgio Milani, Roberta Ramponi,
  Paolo Mataloni, Roberto Osellame, Ernesto~F. Galv{\~a}o, and Fabio Sciarrino.
\newblock Experimental validation of photonic boson sampling.
\newblock {\em Nature Photonics}, 8(8):615--620, Aug 2014.

\bibitem{Spring:17}
Justin~B. Spring, Paolo~L. Mennea, Benjamin~J. Metcalf, Peter~C. Humphreys,
  James~C. Gates, Helen~L. Rogers, Christoph S\"{o}ller, Brian~J. Smith,
  W.~Steven Kolthammer, Peter G.~R. Smith, and Ian~A. Walmsley.
\newblock Chip-based array of near-identical, pure, heralded single-photon
  sources.
\newblock {\em Optica}, 4(1):90--96, Jan 2017.

\bibitem{Faruque:18}
Imad~I. Faruque, Gary~F. Sinclair, Damien Bonneau, John~G. Rarity, and Mark~G.
  Thompson.
\newblock On-chip quantum interference with heralded photons from two
  independent micro-ring resonator sources in silicon photonics.
\newblock {\em Opt. Express}, 26(16):20379--20395, Aug 2018.

\bibitem{doi:10.1116/5.0018594}
S.~Signorini and L.~Pavesi.
\newblock On-chip heralded single photon sources.
\newblock {\em AVS Quantum Science}, 2(4):041701, 2020.

\bibitem{PhysRevLett.113.100502}
A.~P. Lund, A.~Laing, S.~Rahimi-Keshari, T.~Rudolph, J.~L. O'Brien, and T.~C.
  Ralph.
\newblock Boson sampling from a gaussian state.
\newblock {\em Phys. Rev. Lett.}, 113:100502, Sep 2014.

\bibitem{Bentivegnae1400255}
Marco Bentivegna, Nicol{\`o} Spagnolo, Chiara Vitelli, Fulvio Flamini, Niko
  Viggianiello, Ludovico Latmiral, Paolo Mataloni, Daniel~J. Brod, Ernesto~F.
  Galv{\~a}o, Andrea Crespi, Roberta Ramponi, Roberto Osellame, and Fabio
  Sciarrino.
\newblock Experimental scattershot boson sampling.
\newblock {\em Science Advances}, 1(3), 2015.

\bibitem{PhysRevLett.118.020502}
Sonja Barkhofen, Tim~J. Bartley, Linda Sansoni, Regina Kruse, Craig~S.
  Hamilton, Igor Jex, and Christine Silberhorn.
\newblock Driven boson sampling.
\newblock {\em Phys. Rev. Lett.}, 118:020502, Jan 2017.

\bibitem{PhysRevLett.113.120501}
Keith~R. Motes, Alexei Gilchrist, Jonathan~P. Dowling, and Peter~P. Rohde.
\newblock Scalable boson sampling with time-bin encoding using a loop-based
  architecture.
\newblock {\em Phys. Rev. Lett.}, 113:120501, Sep 2014.

\bibitem{PhysRevA.93.043803}
Mihir Pant and Dirk Englund.
\newblock High-dimensional unitary transformations and boson sampling on
  temporal modes using dispersive optics.
\newblock {\em Phys. Rev. A}, 93:043803, Apr 2016.

\bibitem{PhysRevLett.112.050504}
C.~Shen, Z.~Zhang, and L.-M. Duan.
\newblock Scalable implementation of boson sampling with trapped ions.
\newblock {\em Phys. Rev. Lett.}, 112:050504, Feb 2014.

\bibitem{Toyoda2015}
Kenji Toyoda, Ryoto Hiji, Atsushi Noguchi, and Shinji Urabe.
\newblock Hong--ou--mandel interference of two phonons in trapped ions.
\newblock {\em Nature}, 527(7576):74--77, Nov 2015.

\bibitem{PhysRevLett.117.140505}
Borja Peropadre, Gian~Giacomo Guerreschi, Joonsuk Huh, and Al\'an Aspuru-Guzik.
\newblock Proposal for microwave boson sampling.
\newblock {\em Phys. Rev. Lett.}, 117:140505, Sep 2016.

\bibitem{Zhong1460}
Han-Sen Zhong, Hui Wang, Yu-Hao Deng, Ming-Cheng Chen, Li-Chao Peng, Yi-Han
  Luo, Jian Qin, Dian Wu, Xing Ding, Yi~Hu, Peng Hu, Xiao-Yan Yang, Wei-Jun
  Zhang, Hao Li, Yuxuan Li, Xiao Jiang, Lin Gan, Guangwen Yang, Lixing You,
  Zhen Wang, Li~Li, Nai-Le Liu, Chao-Yang Lu, and Jian-Wei Pan.
\newblock Quantum computational advantage using photons.
\newblock {\em Science}, 370(6523):1460--1463, 2020.

\bibitem{zhong2021phase}
Han-Sen Zhong, Yu-Hao Deng, Jian Qin, Hui Wang, Ming-Cheng Chen, Li-Chao Peng,
  Yi-Han Luo, Dian Wu, Si-Qiu Gong, Hao Su, et~al.
\newblock Phase-programmable gaussian boson sampling using stimulated squeezed
  light.
\newblock {\em arXiv preprint arXiv:2106.15534}, 2021.

\bibitem{bulmer2021boundary}
Jacob F.~F. Bulmer, Bryn~A. Bell, Rachel~S. Chadwick, Alex~E. Jones, Diana
  Moise, Alessandro Rigazzi, Jan Thorbecke, Utz-Uwe Haus, Thomas~Van
  Vaerenbergh, Raj~B. Patel, Ian~A. Walmsley, and Anthony Laing.
\newblock The boundary for quantum advantage in gaussian boson sampling, 2021.

\bibitem{DBLP:journals/corr/abs-2009-01177}
Yuxuan Li, Mingcheng Chen, Yaojian Chen, Haitian Lu, Lin Gan, Chao{-}Yang Lu,
  Jian{-}Wei Pan, Haohuan Fu, and Guangwen Yang.
\newblock Benchmarking 50-photon gaussian boson sampling on the sunway
  taihulight.
\newblock {\em CoRR}, abs/2009.01177, 2020.

\bibitem{Gupt2019}
Brajesh Gupt, Josh Izaac, and Nicolás Quesada.
\newblock The walrus: a library for the calculation of hafnians, hermite
  polynomials and gaussian boson sampling.
\newblock {\em Journal of Open Source Software}, 4(44):1705, 2019.

\bibitem{10.5555/898758}
Message~P Forum.
\newblock Mpi: A message-passing interface standard.
\newblock Technical report, USA, 1994.

\bibitem{piquassoboost}
Piquasso boost libraries.
\newblock
  \url{https://github.com/Budapest-Quantum-Computing-Group/piquassoboost},
  2021.

\bibitem{fa13248b3e0a4fa6b7774c0a96d2551a}
{Nicholas J.} Higham.
\newblock Cholesky factorization.
\newblock {\em Wiley Interdisciplinary Reviews: Computational Statistics},
  1(2):251--254, September 2009.

\bibitem{ProTBB}
Michael Voss, Rafael Asenjo, and James Reinders.
\newblock {\em Pro TBB: C++ parallel programming with threading building
  blocks}.
\newblock New York: Apress Open, 2019.

\bibitem{Sutherland1966TheOG}
W.~R. Sutherland.
\newblock The on-line graphical specification of computer procedures.
\newblock 1966.

\bibitem{strawberryfields}
Nathan Killoran, Josh Izaac, Nicol{'{a}}s Quesada, Ville Bergholm, Matthew Amy,
  and Christian Weedbrook.
\newblock {S}trawberry {F}ields: A software platform for photonic quantum
  computing.
\newblock {\em Quantum}, 3:129, 2019.

\end{thebibliography}

\end{document}